\documentclass[12pt]{article}

\catcode`\@=11
\def\@citex[#1]#2{%
\if@filesw \immediate \write \@auxout {\string \citation {#2}}\fi
\@tempcntb\m@ne \let\@h@ld\relax \def\@citea{}%
\@cite{%
  \@for \@citeb:=#2\do {%
    \@ifundefined {b@\@citeb}%
      {\@h@ld\@citea\@tempcntb\m@ne{\bf ?}%
      \@warning {Citation `\@citeb ' on page \thepage \space undefined}}%
      {\@tempcnta\@tempcntb \advance\@tempcnta\@ne%
      \@tempcntb\number\csname b@\@citeb \endcsname \relax%
      \ifnum\@tempcnta=\@tempcntb 
        \ifx\@h@ld\relax%
          \edef \@h@ld{\@citea\csname b@\@citeb\endcsname}%
        \else%
          \edef\@h@ld{\ifmmode{-}\else--\fi\csname b@\@citeb\endcsname}%
        \fi%
      \else
        \@h@ld\@citea\csname b@\@citeb \endcsname%
        \let\@h@ld\relax%
      \fi}%
    \def\@citea{,\penalty\@highpenalty\,}%
  }\@h@ld
}{#1}}

\def\@citeb#1#2{{[#1]\if@tempswa , #2\fi}}
\def\@citeu#1#2{{$^{#1}$\if@tempswa , #2\fi }}
\def\@citep#1#2{{#1\if@tempswa , #2\fi}}

\def\bcites{         
        \catcode`\@=11
        \let\@cite=\@citeb
        \catcode`\@=12
}

\def\upcites{         
        \catcode`\@=11
        \let\@cite=\@citeu
        \catcode`\@=12
}

\def\plaincites{      
        \catcode`\@=11
        \let\@cite=\@citep
        \catcode`\@=12
}

\newcount\hour
\newcount\minute
\newtoks\amorpm
\hour=\time\divide\hour by 60
\minute=\time{\multiply\hour by 60 \global\advance\minute by-\hour}
\edef\standardtime{{\ifnum\hour<12 \global\amorpm={am}%
        \else\global\amorpm={pm}\advance\hour by-12 \fi
        \ifnum\hour=0 \hour=12 \fi
        \number\hour:\ifnum\minute<10 0\fi\number\minute\the\amorpm}}
\edef\militarytime{\number\hour:\ifnum\minute<10 0\fi\number\minute}

\def\draftlabel#1{{\@bsphack\if@filesw {\let\thepage\relax
   \xdef\@gtempa{\write\@auxout{\string
      \newlabel{#1}{{\@currentlabel}{\thepage}}}}}\@gtempa
   \if@nobreak \ifvmode\nobreak\fi\fi\fi\@esphack}
        \gdef\@eqnlabel{#1}}
\def\@eqnlabel{}
\def\@vacuum{}
\def\marginnote#1{}
\def\draftmarginnote#1{\marginpar{\raggedright\scriptsize\tt#1}}
\def\draft{
        \pagestyle{plain}
        \overfullrule=2pt
        \oddsidemargin -.5truein
        \def\@oddhead{\sl \phantom{\today\quad\militarytime} \hfil
        \smash{\Large\sl DRAFT} \hfil \today\quad\militarytime}
        \let\@evenhead\@oddhead
        \let\label=\draftlabel
        \let\marginnote=\draftmarginnote
        \def\ps@empty{\let\@mkboth\@gobbletwo
        \def\@oddfoot{\hfil \smash{\Large\sl DRAFT} \hfil}
        \let\@evenfoot\@oddhead}
        \def\@eqnnum{(\theequation)\rlap{\kern\marginparsep\tt\@eqnlabel}%
        \global\let\@eqnlabel\@vacuum}  }

\def\blackfonts{
        \font\blackboard=msbm10 scaled\magstep1
        \font\blackboards=msbm8
        \font\blackboardss=msbm6
}

\def\prep{         
        \catcode`\@=11
        \input art10.sty
        \catcode`\@=12
        
        \let\small\null
        \def\blackfonts{
                \font\blackboard=msbm10
                \font\blackboards=msbm7
                \font\blackboardss=msbm5
        }
        \let\sl\it
        \twocolumn
        \sloppy
        \voffset=-2.54truecm
        \hoffset=-2.54truecm
        \flushbottom
        \parindent 1em
        \leftmargini 2em
        \leftmarginv .5em
        \leftmarginvi .5em
        \marginparwidth 48pt
        \marginparsep 10pt
        \setlength{\columnsep}{2truecm}
        \setlength{\textwidth}{25.4truecm}
        \setlength{\textheight}{17truecm}
        \baselineskip=16pt
        \oddsidemargin .18truein
        \evensidemargin .17truein
}

\def\eqalign#1{\null\,\vcenter{\openup\jot\m@th
  \ialign{\strut\hfil$\displaystyle{##}$&$\displaystyle{{}##}$\hfil
      \crcr#1\crcr}}\,}
\def\eqalignno#1{\displ@y \tabskip\centering
  \halign to\displaywidth{\hfil$\@lign\displaystyle{##}$\tabskip\z@skip
    &$\@lign\displaystyle{{}##}$\hfil\tabskip\centering
    &\llap{$\@lign##$}\tabskip\z@skip\crcr
    #1\crcr}}

\def\section{\@startsection {section}{1}{\z@}{3.ex plus 1ex minus
 .2ex}{2.ex plus .2ex}{\large\bf}}
\def\subsection{\@startsection{subsection}{2}{\z@}{2.75ex plus 1ex minus
 .2ex}{1.5ex plus .2ex}{\bf}}        

\def\appendix{{\newpage\section*{Appendix}}\let\appendix\section%
        {\setcounter{section}{0}
        \gdef\thesection{\Alph{section}}}\section}

\def\abstract{\if@twocolumn
\section*{Abstract}
\else 
\begin{center}
{\bf Abstract\vspace{-.5em}\vspace{0pt}}
\end{center}
\quotation
\fi}

\catcode`\@=12

\def\sqr#1#2{{\vcenter{\vbox{\hrule height.#2pt\hbox{\vrule width.#2pt 
height#1pt \kern#1pt \vrule width.#2pt}\hrule height.#2pt}}}}

\def\=d{\,{\buildrel\rm def\over =}\,}

\def\i3p{\p32\int d^3p}

\def\As{A\hbox to 1pt{\hss /}}
\def\np4{\int d^4p_1\cdots d^4p_{n-1}\, }

\def\nx4{\int d^4x_1\ldots d^4x_n\, }

\def\kon#1#2{\vbox{\halign{##&&##\cr
\lower4pt\hbox{$\scriptscriptstyle\vert$}\hrulefill &
\hrulefill\lower4pt\hbox{$\scriptscriptstyle\vert$}\cr $#1$&
$#2$\cr}}}

\def\konv#1#2#3{\hbox{\vrule height12pt depth-1pt}
\vbox{\hrule height12pt width#1cm depth-11.6pt}
\hbox{\vrule height6.5pt depth-0.5pt}
\vbox{\hrule height11pt width#2cm depth-10.6pt\kern5pt
      \hrule height6.5pt width#2cm depth-6.1pt}
\hbox{\vrule height12pt depth-1pt}
\vbox{\hrule height6.5pt width#3cm depth-6.1pt}
\hbox{\vrule height6.5pt depth-0.5pt}}
\def\konu#1#2#3{\hbox{\vrule height12pt depth-1pt}
\vbox{\hrule height1pt width#1cm depth-0.6pt}
\hbox{\vrule height12pt depth-6.5pt}
\vbox{\hrule height6pt width#2cm depth-5.6pt\kern5pt
      \hrule height1pt width#2cm depth-0.6pt}
\hbox{\vrule height12pt depth-6.5pt}
\vbox{\hrule height1pt width#3cm depth-0.6pt}
\hbox{\vrule height12pt depth-1pt}}

\def\konw#1#2#3{\hbox{\vrule height12pt depth-1pt}
\vbox{\hrule height12pt width#1cm depth-11.6pt}
\hbox{\vrule height6.5pt depth-0.5pt}
\vbox{\hrule height12pt width#2cm depth-11.6pt \kern5pt
      \hrule height6.5pt width#2cm depth-6.1pt}
\hbox{\vrule height6.5pt depth-0.5pt}
\vbox{\hrule height12pt width#3cm depth-11.6pt}
\hbox{\vrule height12pt depth-1pt}}

\def\i{{\rm int}}

\def\c{{\rm cl}}
\def\e{{\rm ext}}

\def\r{{\rm ret}}
\def\a{{\rm av}}

\def\m3{{\mu_1\mu_2\mu_3}}

\def\p{{(+)}}

\def\be{\begin{equation}}       \def\eq{\begin{equation}}
\def\ee{\end{equation}}         \def\eqe{\end{equation}}

\def\bea{\begin{eqnarray}}      \def\eqa{\begin{eqnarray}}
\def\ena{\end{eqnarray}}        \def\eea{\end{eqnarray}}
                                \def\eqae{\end{eqnarray}}

\def\ba{\begin{array}}
\def\ea{\end{array}}
\def\unit{1 \hskip-.3em \raise2pt\hbox{$ \scriptstyle |$ } }

\def\a{\alpha}
\def\b{\beta}
\def\c{\gamma} 

\def\e{\epsilon}           
\def\f{\phi}               
\def\g{\gamma}
   
\def\i{\iota}

\def\k{\kappa}                    
\def\l{\lambda}
\def\m{\mu}
\def\n{\nu}
  
\def\p{\pi}                
\def\r{\rho}                                     
\def\s{\sigma}                                   
\def\t{\tau}

\def\G{\Gamma}

\def\O{\Omega}

\def\ca{{\cal A}}

\def\cp{{\cal P}}

\def\half{{1 \over 2}}

\def\bop#1{\setbox0=\hbox{$#1M$}\mkern1.5mu
        \vbox{\hrule height0pt depth.04\ht0
        \hbox{\vrule width.04\ht0 height.9\ht0 \kern.9\ht0
        \vrule width.04\ht0}\hrule height.04\ht0}\mkern1.5mu}
\def\pa{\partial}                              

\def\>{\rangle} 

\def\<{\langle} 
\def\Dsl{D \hskip-.6em \raise1pt\hbox{$ / $ } }

\def\sl#1{\rlap{\hbox{$\mskip 1 mu /$}}#1}
\def\leftrightarrowfill{$\mathsurround=0pt \mathord\leftarrow \mkern-6mu
       \cleaders\hbox{$\mkern-2mu \mathord- \mkern-2mu$}\hfill
       \mkern-6mu \mathord\rightarrow$}
\def\dvec#1{\vbox{\ialign{##\crcr
       \leftrightarrowfill\crcr\noalign{\kern-1pt\nointerlineskip}
       $\hfil\displaystyle{#1}\hfil$\crcr}}}          
\def\hook#1{{\vrule height#1pt width0.4pt depth0pt}}
\def\leftrighthookfill#1{$\mathsurround=0pt \mathord\hook#1
       \hrulefill\mathord\hook#1$}
\def\underhook#1{\vtop{\ialign{##\crcr                 
       $\hfil\displaystyle{#1}\hfil$\crcr
       \noalign{\kern-1pt\nointerlineskip\vskip2pt}
       \leftrighthookfill5\crcr}}}
\def\smallunderhook#1{\vtop{\ialign{##\crcr      
       $\hfil\scriptstyle{#1}\hfil$\crcr
       \noalign{\kern-1pt\nointerlineskip\vskip2pt}
       \leftrighthookfill3\crcr}}}

\def\sfrac#1#2{{\vphantom1\smash{\lower.5ex\hbox{\small$#1$}}\over
       \vphantom1\smash{\raise.4ex\hbox{\small$#2$}}}} 
\def\bfrac#1#2{{\vphantom1\smash{\lower.5ex\hbox{$#1$}}\over
       \vphantom1\smash{\raise.3ex\hbox{$#2$}}}}      
\def\afrac#1#2{{\vphantom1\smash{\lower.5ex\hbox{$#1$}}\over#2}}  
\def\on#1#2{{\buildrel{\mkern2.5mu#1\mkern-2.5mu}\over{#2}}}
\def\ddt#1{\on{\hbox{\LARGE .\kern-2pt.}}#1}             
\def\tdt#1{\on{\hbox{\LARGE .\kern-2pt.\kern-2pt.}}#1}   

\def\boxes#1{
       \newcount\num
       \num=1
       \newdimen\downsy
       \downsy=-1.5ex
       \mskip-2.8mu
       \bo
       \loop
       \ifnum\num<#1
       \llap{\raise\num\downsy\hbox{$\bo$}}
       \advance\num by1
       \repeat}
\def\boxup#1#2{\newcount\numup
       \numup=#1
       \advance\numup by-1
       \newdimen\upsy
       \upsy=.75ex
       \mskip2.8mu
       \raise\numup\upsy\hbox{$#2$}}

\newskip\humongous \humongous=0pt plus 1000pt minus 1000pt
\def\caja{\mathsurround=0pt}
\def\eqalign#1{\,\vcenter{\openup2\jot \caja
       \ialign{\strut \hfil$\displaystyle{##}$&$
       \displaystyle{{}##}$\hfil\crcr#1\crcr}}\,}
\newif\ifdtup

\def\to{\rightarrow}

\def\1ov4{{1\over 4}}

\def\pa{\partial}
\def\xx{\times}

\def\ddt{\dot{\t}}

\def\pa{\partial}

\def\xx{\times}

\def\nonu{\nonumber \\{}}
\def\half{{1 \over 2}}

\def\1ad{\mbox{\normalsize $^1$}}
\def\2ad{\mbox{\normalsize $^2$}}
\def\3ad{\mbox{\normalsize $^3$}}
\def\nn{\nonumber \\}

\renewcommand{\theequation}{\thesection.\arabic{equation}}

\begin{document}
 



\null\vskip-24pt
\hfill KUL-TF-99/11
\vskip-10pt
\hfill SPIN-1999/06
\vskip-10pt
\hfill INLO-PUB-7/99
\vskip-10pt
\hfill {\tt hep-th/9904073}
\vskip0.3truecm
\begin{center}
{\Large\bf
AdS/CFT dualities involving large 2d N=4 \\\vskip8pt
superconformal symmetry
}\\ 
\vskip 1.5truecm
{\large\bf
Jan de Boer${}^{*\star}$\footnote{
email:{\tt J.deBoer@phys.uu.nl}}, 
Andrea Pasquinucci${}^\dagger$\footnote{
email:{\tt Andrea.Pasquinucci@cern.ch}} 
and Kostas Skenderis${}^*$\footnote{
email:{\tt K.Skenderis@phys.uu.nl}}
}\\
\vskip 1.5truemm
${}^*$ {\it Spinoza Institute, University of Utrecht,\\
Leuvenlaan 4, 3584 CE Utrecht, The Netherlands}
\vskip 1truemm
${}^\star$ {\it Instituut-Lorentz for Theoretical Physics,
University of Leiden, \\
PO Box 9506, NL-2300 RA, Leiden, The Netherlands}
\vskip 1truemm
${}^\dagger$ {\it Instituut voor Theoretische Fysica, KU Leuven\\
Celestijnenlaan 200D, B-3001 Leuven, Belgium
}\\\vskip 10truemm
\end{center}
\vskip 2truecm
\noindent{\bf Abstract:}
We study the duality between string theory on
$AdS_3 \times S^3 \times S^3$ and two-dimensional conformal theories
with large  $N=4$ superconformal algebra $\ca_\g$. We discuss 
configurations of intersecting branes which give rise to such
near-horizon geometries. We compute the Kaluza-Klein spectrum and
propose that the boundary superconformal theory can be described
by a sigma model on a suitable symmetric product space with a
particular choice of anti-symmetric two-form.
\vfill
\vskip4pt
\noindent{March 1999}

\eject


%
%

%




\section{Introduction}
\setcounter{equation}{0}

Among the AdS/CFT dualities \cite{Malda,Gubs,Wit}
the case of $AdS_3/CFT_2$ is distiguished  because generically one 
has better control of both sides of the correspondence. 
On the one side, one deals with $2d$ superconformal field theories (SCFT) 
which have been quite extensively studied. On the other 
side, one has a three-dimensional $AdS$ gravity which is quite 
simple by itself. In addition, the near-horizon limit 
generically involves (after suitable dualities)
exact WZW models for the various parts that 
make up the near-horizon configuration, so one can 
go beyond the supergravity approximation by considering 
string theory on $AdS_3$\cite{GKS,BORT,KS}.

It is thus quite natural to try to fully explore these cases
and hopefully learn some lessons for the higher dimensional cases as 
well. To that end one can try to formulate the correspondence in 
a setting such that there is as much control over the theory as 
possible. This can happen either if we are dealing with a theory 
with very little structure, like a free fermion, or 
with a theory with a large symmetry group. In this paper 
we will follow the second route. We will study the 
correspondence in the case the boundary superconformal theory 
has the maximal possible linearly realized symmetry, 
i.e. the symmetry algebra is the large (or double) 
$N=4$ superconformal algebra $\ca_\g$ \cite{SeTrPr,IV}. This algebra 
contains two commuting affine $\widehat{SU(2)}$ Lie algebras
(in contrast, the small $N=4$ algebra contains only one affine
$\widehat{SU(2)}$ Lie algebra). It also contains
the finite 
dimensional superalgebra $D^1(2,1,\g/(1-\g))$.
Conformal models with this symmetry algebra are
characterized by two integers, the two levels $k^+$ and $k^-$ 
of the two affine $\widehat{SU(2)}$ algebras 
($\g=k^-/(k^++k^-$)). 

One might expect that these models would be easy to analyze
because of the large amount of symmetry. 
However, this turns out not to be the case as
the structure of the superconformal algebra is quite non-trivial. 
In supersymmetric theories the analysis of BPS sector of the theory is 
usually tractable. For instance, in theories 
with the $N=2$ or the small $N=4$ symmetry algebra there is a linear
relation between 
the conformal dimension and charge of a BPS state.
This implies that the BPS states form a ring. 
In the case of $\ca_\g$ there is still a Bogomolnyi 
bound. The relation, however, between the conformal weight and
the charges is non-linear. This complicates the analysis.
In particular, for the brane realization that we 
will study the non-linear piece is subleading in $1/N$\,\footnote{
$N$ can stand for $k\equiv k^+ +
k^-$, $k^+/k^-$ or $k^-/k^+$. In the large $N$ limit the number
of branes of certain types always becomes large.
When we talk about $1/N$ corrections in later sections
it will be clear from
the context what is meant by $N$.} and
therefore corresponds to string loop corrections. 
Thus, we find the novel situation that  
the mass formula for BPS states receives quantum corrections.
This leads to the possibility that states that 
satisfy a bound in supergravity cease to do so in string theory.
In addition, BPS states 
do not form a ring (but they may form a module over a ring). 

AdS/CFT dual pairs can be obtained by considering 
a configuration of branes and taking a limit in 
which a decoupled worldvolume theory is obtained.
This limit is at the same time a near-horizon limit 
for the corresponding supergravity configuration. The 
near-horizon isometry superalgebra becomes the worldvolume 
superconformal algebra. A brane configuration that leads to 
a dual superconformal theory with symmetry algebra $\ca_\gamma$ 
was identified in \cite{BPS}. It consists of a 
``non-standard'' intersection of two M5 branes with an M2 
brane. In the near-horizon limit the geometry contains 
$AdS_3 \xx S^3 \xx S^3$. The isometry groups of the 
two spheres become the R-symmetry $SU(2)$s, 
and the radii of the spheres are related to the 
levels of the $\widehat{SU(2)}$s. Brane configurations with the
same near-horizon limit exist in all string theories.
In particular, in IIB string theory a configuration
that leads to this geometry consists of 
an overlap of two D1-D5 systems (similarly the M-theory 
configuration may be thought of as an overlap of two 
M2-M5 systems). The near-horizon limit in the case of M-branes
is the standard low-energy limit. In the case of 
IIB branes, however, one needs to consider an ultra-low energy 
limit.

Another motivation for studying this system is that 
it can be viewed as the master configuration 
from which one can reach other systems involving in their 
near-horizon limit a factor of $AdS_2$ or $AdS_3$
and one or two factors of $S^2$ and/or $S^3$ by 
either adding branes and/or taking appropriate 
limits \cite{BPS}. Therefore, a 
complete understanding of the $AdS_3 \xx S^3 \xx S^3$ 
configuration may lead to a unified picture of all 
the other cases as well.

In this paper we begin a detailed analysis of the 
$AdS_3 \xx S^3 \xx S^3/\ca_\g$ duality. 
Previous work can be found in \cite{Giveonetal}.
One of our aims is to identify the dual superconformal theory.
We will propose, generalizing the proposal of \cite{Giveonetal,Giv},
that the dual SCFT 
is a sigma model with target space the symmetric
product Sym${}^{k^-}(U(2))$, 
with $k/k^-$ units of $H=dB$ flux and certain discrete
``gauge fields'' associated to the permutation group
$S_{k^-}$ turned on. The latter are needed to make the
conformal field theory well-defined. An alternative
description is obtained by exchanging the roles of $k^+$
and $k^-$. 

We start our analysis by studying the supergravity solutions.
The first problem one encounters is that the near-horizon
geometry contains an extra non-compact isometry. In order 
to be able to reduce the theory to three dimensions we 
need to compactify this direction. The corresponding 
identifications induce a novel UV/IR relation
between the worldvolume theories of the two overlapping brane configurations. 
Furthermore, the requirement that the Brown-Henneaux central 
charge \cite{BrHe} matches the central charge of the $\ca_\g$
theory
fixes the product of the radius of the extra circle and the 
string coupling constant. The ratio of the two is a modulus of
the solution.
For fixed string coupling constant,
the radius can be made very small at large $N$. 

To obtain some information about the spectrum of the 
conformal field theory we compute the Kaluza-Klein (KK) spectrum 
of supergravity on $adS^3 \times S^3 \times S^3 \times S^1$.
Since the radius of $S^1$ can be made very small in the large 
$N$ limit we start from $9d$ supergravity.
The KK spectrum is obtained by using the group theory method 
developed in \cite{lars,Jan}. The rather tedious computation 
yields a quite simple spectrum. The KK states 
fit in short multiplets of $D^1(2,1,\a)$, but many
short multiplets can be combined into long multiplets and
we get only limited information about the set of BPS states at
a generic point in the moduli space.

Having obtained the single particle KK spectrum the next step is to 
obtain the chiral spectrum of the dual conformal field theory. 
The main question is which of the single and multiparticle 
states correspond to chiral states of the boundary SCFT.
Because of the non-linearity of the BPS bound this is 
a difficult question, especially since the non-linearity 
is invisible in supergravity. We will consider various possible
answers to this question and discuss which ones are
consistent with the above mentioned proposal 
for a candidate boundary SCFT. Notice that to 
verify directly which of the single and multiparticle states are actually 
massive we would need to calculate $1/N$ corrections
as the BPS bound contains terms subleading in $1/N$.

Another way to obtain information about the boundary 
superconformal theory is to use D-brane perturbation 
theory. The boundary theory is then
the infrared limit of the worldvolume 
gauge theory. This is expected to be (a deformation of) 
a sigma model with target space the moduli 
space of vacua of the worldvolume gauge theory.
One may study the D-brane system by introducing a 
probe brane. A new feature in our case is that 
the theory on the probe contains new couplings 
due to loops of open strings that cannot be ignored 
even at low-energies since they involve massless 
particles. These couplings, however, 
are subleading in $1/N$. Thus, only in the large $N$ limit 
one expects that the $\ca_\g$ theory can be described by
a perturbative sigma model. 
Indeed, we find that $1/N$ is what controls the
loop expansion of the
sigma models with $\ca_\g$ symmetry. 

The paper is organized as follows. In  section 2 we present 
the relevant supergravity solutions, both in M-theory 
and in type IIB. The main properties of the large $N=4$
superconformal algebra $\ca_\g$ as well as the closely related non-linear 
algebra $\tilde{\ca}_\g$ are recalled in section 3. Section 4 contains the 
computation of the Kaluza-Klein spectrum. In section 5 we discuss the 
multiparticle spectrum and we present our proposal for the 
boundary SCFT. Section 6 contains some remarks on the  
the D-brane analysis of the system, and section 7 
a discussion of $\s$-models with $\ca_\g$ symmetry.
Open problems are discussed in section 8.
In the appendix we present our conventions.

%
%

\section{Solitonic description}
\setcounter{equation}{0}

In this section we review supergravity solutions which in the 
``near-horizon'' limit yield a solution of the 
form $adS_3 \times S^3 \times S^3 \times S^1$.
Supergravities in 10 dimensions (i.e. type IIA, type IIB and type I)
as well as $11d$ supergravity have solutions describing
intersecting branes that in an appropriate limit 
approach a geometry that contains $adS_3 \xx S^3 \xx S^3$.\footnote{A 
solution of this form
appeared first in the context of the heterotic string theory 
in \cite{ABS}, and as a near-horizon limit
of a configuration of intersecting branes in type I supergravity  
in \cite{CT}.} The type IIA solution can be 
obtained from the M-theory solution by reduction. The type I 
solution can be obtained from either the IIA or the IIB solution.
The M-theory and IIB solution 
seem not to be related by dualities (at least in an obvious manner)
so we will discuss them separately. 

Before presenting the solutions we briefly discuss their 
connection to solutions involving 
a factor of $adS_2$ and/or $S^2$'s in the near-horizon limit.  
There is a simple way, the wave/monopole rule\cite{BPS},
to generate solutions that contain a factor 
of $adS_2$ and/or $S^2$ in their near-horizon limit
starting from a solution that contains $adS_3$ and/or $S^3$. 
To get a factor of $adS_2$ one has to add a wave to the solution, 
after which one has to T-dualize it
(or reduce it if we start from eleven dimensions) 
in the direction of the wave.
To get a solution that involves an $S^2$ one has to add
a KK monopole, after which one has to
T-dualize (or reduce) in the nut-direction.\footnote{
Notice that  the ``near-horizon'' limit 
of the final configuration is a ``very-near-horizon'' limit\cite{stro} of the 
original one. This is so because the T-duality that connects the 
two configuration involves explicit factors of $\a'$.} 
In addition, the wave/monopole rule allows one to determine 
the isometry superalgebra of the new solution
(since one can obtain both the killing spinors and the bosonic isometries
starting from the killing spinors and the bosonic 
isometries of the solution that involves $adS_3$ and/or $S^3$), and 
therefore the symmetry algebra of the dual superconformal theory.
All solutions found in this way involve exact CFT's in their 
near-horizon limit. (Actually one can trace the origin of the 
wave/monopole rule to the relation between the exact CFT's
associated to $adS_3$, $adS_2$ and $S^3$, $S^2$). 
Details can be found in \cite{BPS}. The solutions can be 
further generalized to include rotation \cite{GMT}.
This does not change the near-horizon configuration
as the effect of the rotation can be removed by a
coordinate transformation. Further solutions with the same 
near-horizon geometry can be found in \cite{Pap}.

\subsection{M-theory case}

Consider the configuration
$$
\begin{array}{lccccccccccc}
M5_1 & 1 &   & 3 & 4 & 5 & 6 &   &   &   &    & \\
M5_2 & 1 &   &   &   &   &   & 7 & 8 & 9 & 10 & \\
M2   & 1 & 2 &   &   &   &   &   &   &   &    &
\end{array}
$$
The explicit solution belonging to this configuration 
is given by \cite{TG,Gaunt}
\bea
ds^2 = (H_T)^{\frac{1}{3}}(H_F^{(1)} H_F^{(2)})^{\frac{2}{3}}
 \big\{(H_T H_F^{(1)} H_F^{(2)})^{-1}(-dt^2 +dx_1^2)\nonu
 + (H_T)^{-1} dx_2^2 + (H_F^{(1)})^{-1}(dx_3^2 +\cdots +dx_6^2)
 + (H_F^{(2)})^{-1}(dx_7^2 +\cdots + dx_{10}^2)\big\}\,,
\label{M552}\\
F_{012I}=-\partial_I (H_T)^{-1}\,,\ \ \ F_{2m'n'p'}=\e_{m'n'p'q'}
\partial_{q'} H_F^{(1)}\,,\ \ \ F_{2mnp}=\e_{mnpq}\partial_q H_F^{(2)}
\,,\nonumber
\eea
where $I$ runs over all $m\in\{3,4,5,6\}$ and $m'\in\{7,8,9,10\}$.
$H_F^{(1)}(x')$ and $H_F^{(2)}(x)$ are harmonic functions in the 
relative transverse directions,
\be
H_F^{(1)}=1 + {Q_F^{(1)} \over r'^2}, \qquad H_F^{(2)}=1 + 
{Q_F^{(2)} \over r^2}, 
\ee
where  $r^2 = x_3^2 + \cdots+ x_6^2,\
r'^2 = x_7^2 + \cdots+ x_{10}^2$ and  
$Q_F^{(i)}=N_F^{(i)} l_p^2, i=1,2$. $N_F^{(i)}$ is equal 
(up to a numerical constant) to the number of coincident fivebranes.
$H_T(x,x')$ satisfies \cite{TG,GGPT}
\bea\label{HT}
\left(H_F^{(1)}(x')\partial_x^2 + H_F^{(2)}(x)\partial_{x'}^2
\right)H_T(x,x')=0\, .
\eea
This equation can be solved by 
\be
H_T=(1 + {Q_T^{(1)} \over r'^2}) (1 + {Q_T^{(2)} \over r^2}) 
\ee
The charges $Q_T^{(i)}$ are equal to $N_T^{(i)} l_p^2$, 
where the quantities $N_T^{(1)}$ and $N_T^{(2)}$ are  
membrane densities in $(x^3, x^4, x^5, x^6)$ and 
$(x^7, x^8, x^9, x^{10})$, respectively. 
 Since there are two harmonic functions associated with 
the membrane one may interpret the solution as
an overlap of two M2-M5 systems. In the near horizon
limit the solution will only depend on the product 
$N_T^{(1)} N_T^{(2)}$ which we will denote by $N_T$.

We now consider the low energy limit, in which we keep the
masses of stretched membranes and the lengths in the
$x_2$ direction fixed in Planck units (this means that we keep 
fixed the string coupling constant in the corresponding type IIA
configuration) 
\be
l_p \to 0, \qquad U = {r^2 \over l_p^3}=\mbox{fixed},\qquad
U'={r'^2 \over l_p^3}=\mbox{fixed}.
\ee
The geometry becomes
\be
{ds^2 \over l_p^2}=(Q_3)^{-1}U U' (-dt^2+dx_1^2)+Q_4 dx_2^2+{Q_1 \over 4}
{dU^2 \over U^2} + {Q_2 \over 4} {dU'^2 \over U'^2}
 +Q_1 d\Omega_{(1)}^2 + Q_2 d\Omega^2_{(2)}
\ee
where
\bea
&&Q_1=\left({N_T \over N_F^{(1)}}\right)^{1/3} 
(N_F^{(2)})^{2/3}, \qquad
Q_2=\left({N_T \over N_F^{(2)}}\right)^{1/3} 
(N_F^{(1)})^{2/3}, \nonu
&&Q_3=\left(N_T\right)^{2/3} 
\left(N_F^{(1)} N_F^{(2)}\right)^{1/3}, \qquad
Q_4=\left({N_F^{(1)} N_F^{(2)} \over  N_T}\right)^{2/3}
\eea
We introduce new variables
\be
u^2 = l^2 {U U' \over Q_3}, \qquad 
\l = {l \over 2}\left(\sqrt{{Q_1 \over Q_2}} \log U 
- \sqrt{{Q_2 \over Q_1}} \log U'\right), \qquad 
l=\sqrt{{Q_1 Q_2 \over Q_1 + Q_2}}
\ee
The metric becomes
\be \label{nrM552}
{ds^2 \over l_p^2}=\left[\left({u \over l}\right)^2 (-dt^2 + dx_1^2) 
+ l^2 {du^2 \over u^2}\right] + d \l^2
+Q_1 d\Omega_{(1)}^2 + Q_2 d\Omega^2_{(2)}+Q_4 dx_2^2
\ee
This is a metric for $adS_3 \xx S^3 \xx S^3 \xx E^2$.
The field strengths are equal to
\be
F_{\k\m\n 2}=2 l^{-1} Q_4^{1/2} l_p^3 \e_{\k\m\n} \,,\ \ \ 
F_{\a\b\c 2}=2 N_F^{(2)} l_p^3 \e_{\a\b\c}\,,\ \ \ 
F_{\a'\b'\c' 2}=2 N_F^{(1)} l_p^3 \e_{\a'\b'\c'}\,,
\label{fieldstr}
\ee
where $\k,\mu,\n\in\{0,1,u\}$ are $adS_3$ indices, $\e_{\k\m\n}$ 
is the volume form of the $adS_3$, 
$\a$ and $\a'$ are indices for the two $S^3$ factors, respectively, 
and $\e_{\a\b\c}$ and $ \e_{\a'\b'\c'}$ are volume forms for the
corresponding unit 
spheres. The field strengths are covariantly constant. One can check 
that the  explicit factors of $l_p$ in the solution 
cancel against the factors of $l_p$ in Newton's constant,
so we set $l_p=1$ from now on.

Let us briefly recall the analysis of supersymmetry from \cite{BPS}.
One can easily check that the solution (\ref{M552}) preserves $1/4$ 
of the supersymmetries. In the near-horizon limit this is 
enhanced by a factor of 2. The Killing spinors 
are products of the geometric Killing spinors on $AdS_3$ and the 
three-spheres as we now discuss. To analyze the supersymmetry it is 
most convenient to choose a basis for the Dirac matrices that 
is adapted to the geometry of the space. Such basis is given 
in (18) of \cite{BPS}. The $11d$ spinors $\e$ are also decomposed 
correspondingly: $\e=\eta \otimes \r \otimes \r' \otimes \xi \otimes
\chi$, where $\eta$ is a spinor on $AdS_3$, $\r$ and $\r'$ are
spinors on the two $S^3$'s, $\xi$ is a spinor on the two-dimensional
space spanned by $x_2$ and $\l$, and $\chi$ is an extra two-component 
spinor. All spinors are two-component ones, so we get that $\e$ 
has 32 components as it should. We refer to \cite{BPS} for 
details, here we only give the final solution. The killing spinor
equations are satisfied if $\eta$, $\r$, $\r'$ are geometric
killing spinors on $AdS_3$ and the two $S^3$s:
\bea \label{kill}
&& D_\m \eta \pm \half {Q_4^{1/2} \over l} \g_\m \eta = 0, \nonu
&& D_\a \r \pm {i \over 2} 
{N_F^{(2)} \over Q_1^{3/2}} \g_\a \r = 0, \nonu
&& D_{\a'} \r' \pm {i \over 2} 
{N_F^{(1)} \over Q_2^{3/2}} \g_{\a'} \r' = 0.
\eea
In addition, there is one projection on the $\xi \otimes \chi$ 
spinors, $\cp (\xi \otimes \chi)=\half(1 + \G)(\xi \otimes \chi)=0$,
where 
\be
\G= {l \over Q_4^{1/2}} i\g^2 \g^\l \otimes 
({N_F^{(2)} \over Q_1^{3/2}} \s_3 - {N_F^{(1)} \over Q_2^{3/2}} \s_1)
\ee
where $\g^2, \g^\l$ are gamma matrices in the Euclidean $x_2, \l$ space,
and $\s_1, \s_3$ are Pauli matrices. One can check that $\G$ is 
traceless and $\G^2=1$, so the projection breaks 1/2 of the supersymmetry.   
In (\ref{kill}) the signs are correlated, namely the $11d$ Killing spinor 
is either a product of Killing spinors that solve the 
equations (\ref{kill}) with the plus sign or the ones that 
solve the equations with the minus sign.  Equations (\ref{kill}) 
have maximal number of solutions. Therefore, the near-horizon 
solution preserves 16 supercharges. 
The form of the killing spinors plus the bosonic symmetries 
already imply that the isometry superalgebra 
is $D^1(2,1,\a)$. This has been explicitly verified in \cite{GMT}
by constructing the isometry superalgebra from the killing spinors.

From the M-theory solution (\ref{M552}) we can obtain a solution of 
IIA supergravity describing an intersection of two solitonic fivebranes
over a fundamental string upon reduction over $x_2$. 
Its ``near-horizon'' limit can be obtained from (\ref{nrM552}):
\bea \label{IIasol}
&&ds^2=\left[\left({u \over l}\right)^2 (-dt^2 + dx_1^2) 
+ l^2 {du^2 \over u^2}\right] + Q_4^{1/2} d \l^2
+N_F^{(2)} d\Omega_{(1)}^2 + N_F^{(1)} d\Omega^2_{(2)} \nonu
&&H_{\k\m\n}=2 l^{-1} \e_{\k\m\n} \,,\ \ \ 
H_{\a\b\c}=2 N_F^{(2)} \e_{\a\b\c}\,,\ \ \ 
H_{\a'\b'\c'}=2 N_F^{(1)} \e_{\a'\b'\c'}\,,  \nonu
&&e^{-2 \f} = {N_T\over N_F^{(1)}N_F^{(2)}}
\eea
where 
\be
{1 \over l^2} = {1 \over N_F^{(1)}} + {1 \over N_F^{(2)}}
\ee
and\footnote{ 
We use the convention to leave a factor of $g_s^2$ in Newton's
constant, so in the full solution the dilaton field vanishes asymptotically. 
S-duality acts as $g_s \to 1/g_s, \a' \to \a' g_s$,
and T-duality as $R \to \a'/R, g_s \to g_s \sqrt{\a'}/R$,
see \cite{rev} for details.} $g_s=(R_2)^{3/2}$, where $R_2$ is the 
radius of $x_2$.\footnote{Notice that $N_T^{(i)}, i=1,2$ are the membrane
densities in eleven dimensional Planck units. Since the $11d$ Planck 
scale differs from the string scale by factors of $g_s$, the IIA 
solution that describes an intersection of a number of solitonic fivebranes 
with a density (in $10d$ units) of fundamental strings  will involve  
extra factors of $g_s$ in the dilaton field.}

Notice that the values of the fields in (\ref{IIasol}) are the canonical ones
such that there are exact CFT's associated with each factor, namely 
an $SL(2, R)$ WZW model for the $adS_3$ part and two $SU(2)$ WZW 
models at level $N_F^{(1)}$ and $N_F^{(2)}$, respectively,
for the two $S^3$'s. This implies a quantization condition for 
$N_F^{(i)}, i=1,2$. In the original solution this quantization was due to a 
quantization of the magnetic fluxes over the two $S^3$'s.

\subsection{IIB case}

The type IIB configuration that has a ``near-horizon'' limit of the form
$adS_3 \xx S^3 \xx S^3$ can be thought of as an overlap of two 
D1-D5 systems,
$$
\begin{array}{lcccccccccc}
D5_1   & 1 & 2 & 3 & 4 & 5 &   &   &   &   & \\
D1_1   & 1 &   &   &   &   &   &   &   &   & \\
D5_2   & 1 &   &   &   &   & 6 & 7 & 8 & 9 & \\
D1_2   & 1 &   &   &   &   &   &   &   &   &
\end{array}
$$
The harmonic functions of the first D1-D5 system depend on the 
relative transverse directions of the second D1-D5 system and 
vice versa. The explicit form of the solution is
\bea \label{sol1}
ds^2&=&(H^{(1)}_1 H^{(1)}_5)^{-1/2} (H^{(2)}_1 H^{(2)}_5)^{-1/2}
(-dt^2 + dx_1^2) \\
&+&\left({H_1^{(1)} \over H_5^{(1)}}\right)^{1/2} (H^{(2)}_1 H^{(2)}_5)^{1/2}
(dr^2 + r^2 d\O_{(1)}^2) 
+\left({H_1^{(2)} \over H_5^{(2)}}\right)^{1/2} (H^{(1)}_1 H^{(1)}_5)^{1/2}
(dr'^2 + r'^2 d\O_{(2)}^2) \nn
H_{01I}&=&-\pa_I(H_1^{(1)} H_1^{(2)})^{-1}, 
\qquad H_{m'n'p'} = \e_{m' n' p' q'} \pa_{q'} H_5^{(1)} 
\qquad H_{mnp} = \e_{m n p q} \pa_{q} H_5^{(2)}  \nn
e^{-2 \f}&=& \left({H_5^{(1)} \over H_1^{(1)}}\right)
\left({H_5^{(2)} \over H_1^{(2)}}\right), \qquad
r^2=x_2^2 + \cdots + x_5^2, \qquad r'^2=x_6^2 + \cdots + x_9^2 \nonumber
\eea
where $I$ runs over all $m\in\{2,3,4,5\}$ and $m'\in\{6,7,8,9\}$.
The harmonic functions are equal to
\bea
&&H_1^{(1)}=1+{Q_1^{(1)} \a' \over r'^2}, \qquad 
H_5^{(1)}=1+{Q^{(1)}_5 \a' \over r'^2}
\nn
&&H_1^{(2)}=1+{Q_1^{(2)} \a' \over r^2}, \qquad 
H_5^{(2)}=1+{Q_5^{(2)} \a' \over r^2}
\eea
where $Q_{k}^{(i)}=g_s N_k^{(i)}, k=1,5, i=1,2$. $N_5^{(i)}, i=1,2$
are (up to a numerical constant) the number of D5 branes. $N_1^{(i)}, i=1,2$,
are D1 brane densities. As in the M-theory solution, in the near
horizon limit the solution only depends on the product 
$N_1^{(1)} N_1^{(2)}$ which we will denote by $N_1$. 

Let us discuss the low-energy limit. We consider the limit
\be 
\a' \to 0,\ U={r \over \a'}=\mbox{fixed},\ U'={r' \over \a'}=\mbox{fixed},\
\tilde{t}={t \over \sqrt{\a'}}=\mbox{fixed},\ 
\tilde{x}_1={x_1 \over \sqrt{\a'}}=\mbox{fixed}.
\ee
The scaling of $x_1, t$ is necessary so that
all terms in the metric get an overall factor of $\a'$.
This limit can be interpreted as an ultra-low energy limit.

The metric becomes
\be
{ds^2 \over \a'}=Q_3^{-1} U^2 U'^2 (-d\tilde{t}^2 + d\tilde{x}_1^2)+
Q_1 {dU^2 \over U^2} + Q_2 {dU'^2 \over U'^2} + 
+Q_1 d\Omega_{(1)}^2 + Q_2 d\Omega^2_{(2)}
\ee
where 
\be \label{Qdefs}
Q_1 = g_s \sqrt{{N_1 N_5^{(2)} \over N_5^{(1)}}}, \qquad  
Q_2 = g_s \sqrt{{N_1 N_5^{(1)} \over N_5^{(2)}}}, \qquad 
Q_3 = g_s^2  \sqrt{N_1 N_5^{(1)} N_5^{(2)}}. 
\ee

This metric is of the form $adS_3 \xx S^3 \xx S^3 \xx R$. To see this
we further change variables
\be
u={l \over \sqrt{Q_3}} U U', \qquad 
\l = l\left(\sqrt{{Q_1 \over Q_2}} \log U 
- \sqrt{{Q_2 \over Q_1}} \log U'\right), \qquad 
l=\sqrt{{Q_1 Q_2 \over Q_1 + Q_2}}.
\ee
The solution takes the form (we drop the tilde from $t, x_1$)
\bea \label{metric}
&&{ds^2 \over \a'}= \left[\left({u \over l}\right)^2 (-dt^2 + dx_1^2) 
+ l^2 {du^2 \over u^2}\right] + d \l^2 + Q_1 d \O_{(1)}^2 + Q_2 d \O_{(2)}^2 
\\
&&H_{\k \m \n}=2 l^{-1} {Q_3 \over Q_1 Q_2} \a' \e_{\k \m \n}, \qquad
H_{\a \b \g} = 2 g_s N_5^{(2)} \a' \e_{\a \b \g}, \qquad
H_{\a' \b' \g'}= 2 g_s N_5^{(1)} \a' \e_{\a' \b' \g'} \nn
&&e^\f = {Q_1 Q_2 \over Q_3} \nonumber
\eea 
where $\k, \m, \n=0,1,u$, are $adS_3$ indices, $\e_{\k \m \n}$
is the volume form of $adS_3$, $\a$ and $\a'$ 
are indices on the two spheres, and $\e_{\a \b \g},
\e_{\a' \b' \g'}$ are the corresponding unit volume forms.
All 
factors of $\a'$ cancel at the end, so we set $\a'=1$ from 
now on.

The metric in (\ref{metric}) is of the form 
$adS_3 \times S^3 \times S^3 \times R$. The $adS_3$ radius is equal to 
$l$ and the radii of the two spheres are equal to $Q_1^{1/2}$ and 
$Q_2^{1/2}$, respectively.
We would like to compactify the $\l$ coordinate, but leave 
the rest of the configuration intact. 
A similar discussion applies to the M-theory configuration in
section~2.1.
This can be achieved by 
the following identification
\be \label{identif}
U \sim U e^{L}, \qquad
U' \sim U' e^{-L}
\ee
This identification leaves invariant the $adS_3 \times S^3 \times S^3$ 
part of the metric and implies
\be
\l \sim \l + L (Q_1 + Q_2)^{1/2}
\ee
Since $U$ and $U'$ can be thought of as a cut-off energy of the 
two D1-D5 systems, the identification (\ref{identif}) imposes some kind of UV-IR
identification between the two D1-D5 systems. 
At this stage the parameter $L$ 
seems a free parameter. However, its value is fixed to a specific value
in order to get the correct central charge for the boundary conformal 
field theory. 

The value of the central charge of the boundary conformal field theory 
is given by\cite{BrHe,HS}
\be
c= {3 l \over 2 G_N^{(3)}}
\ee
In our case we get (ignoring all numerical factors)
\be
c \sim g_s^{-2} Q_3^2 L = g_s^2 N_1 N_5^{(1)} N_5^{(2)} L  .
\ee 
The level $k$ of a current algebra originating from the isometries
of the internal space is proportional to the square of the radius
of that space; this can be seen either via KK reduction or 
directly in string theory. In particular,
the level $k$ of the $SU(2)$ current algebra associated to a three-sphere
of radius $R$ is given by
\be k={R^2 \over 4 l G_N^{(3)}}.
\ee
Therefore, $k^{\pm}$ are given by
\be k^+ = {c \over 6} ( 1 + {N_5^{(2)} \over N_5^{(1)} }),\quad
k^- = {c \over 6} ( 1 + {N_5^{(1)} \over N_5^{(2)} }) .
\ee
From this we deduce that $k^+/k^-=N_5^{(2)}/N_5^{(1)}$. 
This is consistent with the values $k^+ = N_1 N_5^{(2)}$ and 
$k^- = N_1 N_5^{(1)}$ that one obtains by requiring that  
$k^+$ and $k^-$ 
are the levels of the affine $\widehat{SU(2)}$ Lie algebra 
of the dual conformal field theory for each separate D1-D5 system. 
The central charge of the $\ca_{\gamma}$ algebra is equal to 
$c=6 k^+ k^- /(k^+ + k^-)$. Therefore,
\be \label{Lid}
L \sim {1 \over g_s^2 (k^+ + k^-)} 
{k^+ k^-\over N_1 N_5^{(1)} N_5^{(2)} }
\ee
In the large $k^+$ (or $k^-$) limit one of the D5 systems decouples.
The central charge goes over to $c=6 k^-$, and $g_s^2 
L \to 0$ so there are no identifications, as it should. 

The S-dual configuration consists of an intersection of two IIB NS5-F1 
systems. The metric, antisymmetric tensor and dilaton are exactly 
as in the IIA configuration described in the previous section.

%
%


\section{The $\ca_\gamma$ and $\tilde{\ca}_\gamma$ SCFT}
\setcounter{equation}{0}

In this section we review the $N=4$ double SCFTs and discuss some of 
their properties. The $N=4$ double SCFTs are based on the
one-parameter family of $N=4$ super conformal algebras (SCA)
$\ca_\gamma$. These algebras contain, among other sub-structures, two
commuting $\widehat{SU(2)}$ affine Lie algebras and a finite
sub-superalgebra $D(2,1,\gamma/(1-\gamma))$, where $\gamma \in$R. They
are parametrized by the central charge of the Virasoro algebra $c$ and
the parameter $\gamma$ or equivalently by the levels $k^+$ and $k^-$
of $\widehat{SU(2)}^+$ and
$\widehat{SU(2)}^-$. The parameters are related by
\be
c= {6 k^+k^-\over k} \qquad\qquad {\rm with}\quad k\equiv k^+ + k^-=
{c \over 6 \gamma (1-\gamma)}
\ee
Besides the Virasoro generators, the algebra is generated by the two
sets of $\widehat{SU(2)}^\pm$ generators $A^{\pm i}$, a
$\widehat{U(1)}$ generator $U$, all with dimensions 1; four
supersymmetry generators $G^a$ with dimensions $3/2$ and four
fermionic generators $Q^a$ with dimensions $1/2$. 
The operator product expansions of these fields are given in
Appendix A.

We are interested in the unitary representations of this algebra
\cite{Gunay,PT1,PT2},
where $k^+$ and $k^-$ are positive integers, and in particular in the
primary states in the Neveu-Schwarz (NS) sector. 
We thus consider the unitary highest weight states
$\vert hws\rangle$ defined by~\footnote{By convention, the $SU(2)$ indices $i,j$
run over $1,2,3$ or $+,-,3$, and the indices $a,b$ over $1,2,3,4$ or
$+,-,+K,-K$; see ref.\ \cite{Gunay} for more details.}
\bea \label{hwsdef}
&& (L)_n \vert hws\rangle = (A^{\pm i})_n \vert hws\rangle = (U)_n
\vert hws\rangle = (A^{\pm +})_0 \vert hws\rangle = 0 \ {\rm
for}\ n=1,2,\dots,\ i=1,2,3 \nn
&& (G^a)_r \vert hws\rangle = (Q^a)_r \vert hws\rangle =0 \qquad {\rm
for} \quad r={1\over 2},{3\over 2},\dots,\ a=1,2,3,4 
\eea
Thus each $hws$ is characterized by its conformal dimension $h$,
$U(1)$ charge $u$, and the spins $l^\pm$ under the two $SU(2)$s.
The unitarity of the representation implies that
\bea
&& l^\pm =0, {1\over 2},1,\dots, {1\over 2}(k^\pm -1)
\nonumber  \\ 
&& k h\geq (l^+-l^-)^2 + k^- l^+ + k^+ l^- + u^2 
\eea
In supersymmetric conformal field theories based on the $N=2$
or small $N=4$ algebra, the states 
which saturate
some bound for the conformal dimension, are in short 
multiplets and form a ring, called ``chiral ring''. 
Indeed also for the double $N=4$ algebras the $hws$ for which 
$k h = (l^+-l^-)^2 + k^- l^+ + k^+ l^- + u^2$ satisfy the chirality 
(or massless) condition
\be \label{chiraldef}
(\tilde{G}^+)_{-1/2} \vert hws\rangle =0
\ee
where $\tilde{G}$ is defined in eq.\ (\ref{tildedef}). It is important
to notice that in the case of the $\ca_\gamma$ algebra, the ``chiral'' states
do not form any obvious ring, due to the quadratic dependence on the $U(1)$
charge and the $SU(2)$ spins. We will come back to this issue 
after having introduced the $\tilde{\ca}_\gamma$ algebra. 

The $\ca_\gamma$ algebra is related \cite{GoSc} to another SCA called
$\tilde{\ca}_\gamma$, or non-linear $\ca_\gamma$, with no dimension $1/2$
generators, but composite operators in the operator product expansion
(OPE). We will denote all operators belonging to this algebra by a
tilde. Indeed, starting from the $\ca_\gamma$ algebra, if one introduces
the operators~\footnote{See Appendix A  for the 
definition of the $\alpha$ and $\epsilon$ symbols.}
\bea \label{tildedef}
&& \tilde{L}= L+{1\over k} (UU +\partial Q^a Q_a) \\
&&\tilde{G}_a = G_a +{2\over k} UQ_a -{2\over 3k^2} \epsilon_{abcd}
Q^bQ^cQ^d +{4\over k} Q^b(\alpha^{+i}_{ba} \tilde{A}^+_i -
\alpha^{-i}_{ba} \tilde{A}^-_i)\nn
&&\tilde{A}^{\pm i} = A^{\pm i} -{1\over k} \alpha^{\pm i}_{ab}
Q^aQ^b\nn 
&&\tilde{Q} = Q \qquad\qquad\qquad\qquad\qquad\qquad \tilde{U}=U \nonumber
\eea
one can show that the $\tilde{U}$ and $\tilde{Q}$ operators decouple 
completely from the
others and form a free algebra. The other operators form a SCA with
central charge $\tilde{c}= c-3$ with respect to which $\tilde{G}$ and
$\tilde{A}$ are primary operators with dimension $3/2$ and $1$
respectively. The $\tilde{A}^\pm$ are $\widehat{SU(2)}^\pm$ 
affine Lie algebra generators 
with levels 
\be
\tilde{k}^\pm = k^\pm -1
\ee
The unitary $hws$ representations of the two algebras are in
``one-to-one'' correspondence. Indeed to each representation of $\ca_\gamma$
with $h,l^\pm,u$ there corresponds a representation of 
$\tilde{\ca}_\gamma$ 
with 
\be
\tilde{h}= h- {u^2\over k}\quad ,\qquad\qquad \tilde{l}^\pm = l^\pm
\ee
Inversely, to each representation of $\tilde{\ca}_\gamma$ there
corresponds an infinite set of representations of $\ca_\gamma$ built by
adding one free boson and four free fermions and choosing some $u$ 
compatible with the radius of the $\widehat{U(1)}$ generator $U$.

To compare with the spectrum that we obtain from the supergravity
analysis, we are interested in the explicit form of the ``chiral'' 
multiplets in the $\tilde{\ca}_\gamma$ algebra. The $hws$ states are
now defined by 
\bea \label{hwstdef}
&& (\tilde{L})_n \vert hws\rangle = (\tilde{A}^{\pm i})_n \vert
hws\rangle = (\tilde{A}^{\pm +})_0 \vert hws\rangle = 0 \ {\rm
for}\ n=1,2,\dots,\ i=1,2,3 \nn
&& (\tilde{G}^a)_r \vert hws\rangle =0 \qquad {\rm
for} \quad r={1\over 2},{3\over 2},\dots,\ a=1,2,3,4 
\eea
The ``chiral'' condition is again given by eq.\ (\ref{chiraldef}). 
The conformal dimension of a ``chiral'' state is now 
\be \label{eq3.9}
k \tilde{h}_c = (l^+ - l^-)^2 + k^- l^+ + k^+ l^- 
\ee
The explicit form of the ``chiral'' multiplet is the same 
as the one of the $D^1(2,1,\a)$ given in (\ref{auxxx})-(\ref{lowsp}).
As in the case of the linear algebra, these states do not form a
ring. However, since
\be \label{eq3.10}
k \tilde{h}_c(l^+,l^-) + k \tilde{h}_c(m^+,m^-) = 
k \tilde{h}_c(l^+ + m^+ ,l^- + m^-) - 2(l^+ - l^-)(m^+ - m^-)
\ee
if either $l^+=l^-$ or $m^+ = m^-$, the OPE of the two ``chiral''
operators corresponding to these states contains no singular
terms and the products of these two operators can give rise to another
``chiral'' operator. This shows that in contrast to the case of
the usual $N=2$ or $N=4$ algebra, the chiral operators do not
form a ring but they could form a module over a ring.
The latter ring is generated by all ``chiral'' fields of the
form $l^+=l^-$. 

We conclude this section by discussing some general constraints
on the spectrum of chiral operators in any $\tilde{\ca}_{\gamma}$
theory that were derived in \cite{Oogurin4}.
Let $N_{l^{\pm},\bar{l}^{\pm}}$ be the
number of chiral operators with left- and right-moving 
$SU(2)$ quantum numbers
$(l^+,l^-;\bar{l}^+,\bar{l}^-)$.
Consider the generating
function
\be \label{deff}
f(p,q,r,s)=\sum_{l^{\pm},\bar{l}^{\pm}} N_{l^{\pm},\bar{l}^{\pm}}
p^{2 l^+} q^{2 l^-} r^{2 \bar{l}^+} s^{2 \bar{l}^-} 
\ee
According to \cite{Oogurin4} certain linear combinations of the
multiplicities $N_{l^{\pm},\bar{l}^{\pm}}$ should give rise
to $SU(2)$ modular invariants. These linear combinations
are contained in another generating function defined by
 \be \label{defg}
g(x,y) = \sum_{l,\bar{l}=0}^{(\tilde{k}^+ + \tilde{k}^-)/2}
x^{2l} y^{2\bar{l}} \sum_{l^{\pm},\bar{l}^{\pm}=0}^{\tilde{k}^{\pm}/2}
(-1)^{2l^-+2\bar{l}^-} 
\delta_{\tilde{k}^- +  2l^+ -  2l^-,2l}
\delta_{\tilde{k}^- + 2\bar{l}^+ - 2\bar{l}^-,2\bar{l}}
N_{l^{\pm},\bar{l}^{\pm}}
\ee
It is easy to see that
\be \label{gprop}
g(x,y)\equiv \sum_{l,\bar{l}=0}^{(\tilde{k}^+ + \tilde{k}^-)/2}
x^{2l} y^{2\bar{l}}n_{l,\bar{l}}  = (xy)^{\tilde{k}^-} f(x,-x^{-1},y,-y^{-1}).
\ee
The numbers $n_{l,\bar{l}}$ should correspond 
to a modular invariant of $SU(2)$. For example,
the diagonal modular invariant appears when
the function $f$ satisfies 
\be \label{e1} 
f(x,-x^{-1},y,-y^{-1}) = A( (xy)^{-\tilde{k}^-} + 
(xy)^{1-\tilde{k}^-} + \ldots + 
(xy)^{\tilde{k}^+-1} + (xy)^{\tilde{k}^+} )
\ee
for some integer $A$.

%
%

\section{Kaluza-Klein Spectrum}
\setcounter{equation}{0}

To compute the Kaluza-Klein Spectrum of $10d$ supergravity 
on $adS_3 \xx S^3 \xx S^3 \xx S^1$
we use representation
theory and follow the method explained in \cite{Jan}. 
In other words, we will only compute the quantum numbers of the KK modes
under the $SO(4)\times SO(4)$ isometry group of $S^3 \times S^3$,
and deduce the conformal weights of the KK states from that.
In the large $k$ limit (and with $g_s$ fixed) the radius of 
$S^1$ is small so we start from nine dimensional 
supergravity. 

The first step in this procedure is to determine the relevant $AdS$
supergroup, which is a symmetry of nine-dimensional supergravity on
$AdS_3 \times S^3 \times S^3$. KK modes have
to fall into multiplets of this $AdS$ supergroup. As
we discussed in section~2, in the case at
hand the relevant supergroup is $D^1(2,1,\alpha)\times 
D^1(2,1,\alpha)$\footnote{Recall that $\alpha=\gamma/(1-\gamma)=k^+/k^-$}
\cite{BPS}. Indeed, in general the chiral algebra of the
boundary CFT is the Hamiltonian reduction of the affine Lie
superalgebra made from the $AdS$ supergroup, and the 
$\tilde{\ca}_{\gamma}$ SCA can be obtained via Hamiltonian
reduction from $\widehat{D^1(2,1,\alpha)}$ \cite{Bowcock}. 
Also notice that the
bosonic subalgebra of $D^1(2,1,\alpha)\times D^1(2,1,\alpha)$ is
$SU(2)^4 \times SL(2,R)^2$, which is the isometry group of
$AdS_3 \times S^3 \times S^3$. 

The next step is to study the representations of $D^1(2,1,\alpha)$.
We can think of $D^1(2,1,\alpha)$ as being generated by the 
generators $L_{\pm 1},L_0,G^a_{\pm 1/2},A^{\pm i}_0$
of the $\ca_{\gamma}$ algebra. The representation theory of
$D^1(2,1,\alpha)$ mirrors that of $\ca_{\gamma}$. There are
long and short representations. Representations are labeled
by the $SU(2)$ quantum numbers $l^{\pm}$ and the conformal
weight $h$, and will be denoted by $(l^+,l^-,h)$. Unitarity
implies the bound $h \geq \gamma l^- + (1-\gamma)l^+$.
When this bound is saturated 
the corresponding representation is a short
representation, which will be denoted by $(l^+,l^-)_S$.
Each $D^1(2,1,\alpha)$ representation can be decomposed in
terms of representations of the $SU(2)\times SU(2)$ 
subgroup whose representations will be labeled by $(j,j')$.
Generic short representations contain $8$ $SU(2)\times SU(2)$
representations, namely
\bea \label{auxxx}
(l^+,l^-)_S & \rightarrow & 2(l^+,l^-) + 2(l^+-\frac{1}{2},l^--\frac{1}{2} ) 
 + (l^+-\frac{1}{2},l^-+\frac{1}{2}) 
\nonumber \\ & & + (l^++\frac{1}{2},l^--\frac{1}{2}) + (l^+-1,l^-) + (l^+,l^--1).
\eea
These $8$ $SU(2)\times SU(2)$ representations can be organized 
according to the action of $G^a_{-1/2}$ in
the following way
\bea 
& (l^+,l^-) & 
\nonumber \\
& (l^+-\frac12,l^--\frac12 )  \qquad  (l^++{1\over2},l^--{1\over2}) \qquad
 (l^+-\frac12,l^-+\frac12)&
\nonumber \\ 
&(l^+,l^--1) \qquad (l^+-1,l^-)\qquad (l^+,l^-) &
\nonumber \\ 
& (l^+-\frac12,l^--\frac12) . & 
\eea
The states on the first line have conformal weight $h = \gamma l^- + (1-\gamma) l^+$,
and the conformal weight increases by $\frac12$ as we move down one line. 
This result (\ref{auxxx}) 
is only valid if $l^{\pm} \geq 1$. For other values of $l^{\pm}$
the decomposition reads
\bea \label{lowsp}
(\frac{1}{2} ,l^-)_S & \equiv & 2(\frac{1}{2},l^-) + 2(0,l^--\frac{1}{2}) + 
(0,l^-+\frac{1}{2})+(1,l^--\frac{1}{2}) + (\frac{1}{2},l^--1) \nonumber \\
(0,l^-)_S & \equiv & (0,l^-) + (\frac{1}{2},l^--\frac{1}{2}) + (0,l^--1) \nonumber \\
(\frac{1}{2},\frac{1}{2})_S & \equiv & 2(\frac{1}{2},\frac{1}{2}) + 2(0,0) + 
(0,1) + (1,0) \nonumber \\
(0,\frac{1}{2})_S & \equiv & (0,\frac{1}{2}) + (\frac{1}{2},0) \nonumber \\
(0,0)_S & \equiv & (0,0) 
\eea
and similarly for $l^+ \leftrightarrow l^-$.

The KK spectrum can now be determined as in \cite{Jan}. We start
with nine-dimensional supergravity, determine using representation
theory the $SU(2)^4$ quantum numbers of all KK states, and organize
those quantum numbers in terms of short representations of  $D^1(2,1,\alpha)
\times D^1(2,1,\alpha)$. Short representations of 
$D^1(2,1,\alpha) \times D^1(2,1,\alpha)$ are simply the tensor product
of two short representations $(l^+,l^-)_S$ and $(\bar{l}^+,\bar{l}^-)_S$,
and will be denoted by $(l^+,l^-;\bar{l}^+,\bar{l}^-)_S$. 
Each of these contains $64$ $SU(2)^4$ representations, and it is a non-trivial
check on the correctness of the set of $SU(2)^4$ quantum numbers of the
KK states to see if they organize in appropriate groups of $64$.
Once we organize the KK states in short representations we also 
know their conformal weights. The reason that we expect only
short representations to appear in the KK spectrum is that all
KK fields originated from massless fields in nine-dimensions.
Thus, they saturate the inequality $m^2 \geq 0$ and it is natural
to identify this with the bound on the conformal weight of a 
$D^1(2,1,\alpha)$ representation. 

We omit the details of the calculation, but 
it turns out that the KK spectrum can
indeed be organized in terms of short representations and 
the final result for the KK spectrum reads
\bea \label{KKspec}
&&  \oplus_{l^+\geq0,l^-\geq1/2} (l^+,l^-;l^+,l^-)_S +
\oplus_{l^+\geq1/2,l^-\geq0} (l^+,l^-;l^+,l^-)_S + \\
&& \oplus_{l^+,l^-\geq0} \left((l^+,l^-;l^++\frac12,l^-+\frac12)_S +
(l^++\frac12,l^-+\frac12;l^+,l^-)_S \right)\nonumber
\eea
An important remark is that the highest weight states with quantum
numbers 
 $(l^+,l^-;l^+,l^-)_S$ are bosonic, whereas 
$(l^+,l^-;l^++\frac12,l^-+\frac12)_S$ and 
$(l^++\frac12,l^-+\frac12;l^+,l^-)_S$ are fermionic.

We included in (\ref{KKspec})
two short representations that do not correspond
to propagating degrees of freedom in the bulk but to
nontrivial fields on the boundary. These are the short multiplets
$(0,0;\frac12,\frac12)_S$ and $(\frac12,\frac12;0,0)_S$, which
contain all higher modes of the $\ca_{\gamma}$ algebra, in particular
the stress-energy tensor of the boundary theory.  They arise via
a suitable supersymmetric generalization of \cite{BrHe}.

%
%

\section{Multiparticle spectrum and boundary SCFT}
\setcounter{equation}{0}

In the previous section we determined the KK spectrum of single particle
states in supergravity, see (\ref{KKspec}). We organized the spectrum in
terms of short representations of $D^1(2,1,\alpha)\times D^1(2,1,\alpha)$.
The first issue we want to address in this section 
is which of the single and multiparticle states correspond to chiral operators
of the boundary SCFT. Because of the nonlinearity of the bound (\ref{eq3.9})
this is a rather difficult question, especially since the nonlinear part
of (\ref{eq3.9}) is invisible in supergravity where $k^{\pm}$ are very large.
A priori there are at least four options:\\
\noindent
(i) all multiparticle states correspond to chiral operators of the boundary 
SCFT.\\
\noindent
(ii) only products of one single particle state with arbitrary many states 
of the
form $(l,l;\bar{l},\bar{l})_S$ are massless; this is inspired by equation
(\ref{eq3.10}) and the discussion below it, in which we argued the most
natural structure on the space of chiral operators is that of a module over
a ring.\\
\noindent
(iii) Only the single particle states correspond to
chiral operators of the boundary SCFT.\\
\noindent
(iv) Except for powers of the representations 
$(0,0;\frac12,\frac12)_S$,
$(\frac12,\frac12;0,0)_S$ and
$(\frac12,\frac12;\frac12,\frac12)_S$, there are no multiparticle
states corresponding to chiral operators. The motivation for this is
that long representations of $D^1(2,1,\alpha)$, as far as their
$SU(2)\times SU(2)$ quantum numbers go, contain two short
$D^1(2,1,\alpha)$ representations
\be
(l^+,l^-)_{\rm long} = (l^+,l^-)_S + (l^++\frac12,l^-+\frac12)_S .
\ee
From this we see that the entire KK spectrum (\ref{KKspec}) can
be written as the sum of $(0,0;\frac12,\frac12)_S$,
$(\frac12,\frac12;0,0)_S$ and
$(\frac12,\frac12;\frac12,\frac12)_S$,
and $({\rm long}) \otimes ({\rm long})$ representations. The 
conformal weight of these long representations is not protected
by anything and they can therefore correspond to nonchiral
operators. In particular the spectrum of chiral operators can
jump as we move around in the moduli space. 

Unfortunately, there only a few things we can do to decide whether
one of the options (i)-(iv) gives the right spectrum of chiral
operators. One possibility is to examine whether the resulting
spectrum is in agreement with the modular invariance constraints
discussed at the end of section~3. Another possibility is to compute
the spectrum of chiral operators of known $\ca_{\gamma}$ theories
and to match those to the KK spectrum in the hope of identifying
the precise boundary theory. The first possibility is somewhat 
problematic to apply. First of all, we do not know where to truncate
the spectrum, and furthermore, there is no obvious map from a
unitary $\ca_\gamma$ theory to a unitary $\tilde{\ca}_{\gamma}$
with finite multiplicities. Still, if we assume that the
spectrum of chiral primaries of an $\ca_{\gamma}$
theory should satisfy the same constraints as the spectrum of
an $\tilde{\ca}_{\gamma}$ and if we truncate the spectrum to
keep only states with  
$l^{\pm}_{\rm total}< k^{\pm}/2$ and $\bar{l}^{\pm}_{\rm total}
< k^{\pm}/2$ we see that option (i) is not compatible
with modular invariance, but, quite 
surprisingly,  options
(ii), (iii) and (iv) are compatible with modular
invariance. Nevertheless, we should probably not attach
too much value to these observations.

What about the spectrum of known $\ca_{\gamma}$ theories? 
Almost all known $\ca_{\gamma}$ and $\tilde{\ca}_{\gamma}$
theories are associated to certain so-called Wolf spaces
\cite{SeTrPr,SeTrPrSp,Pr,SeTh}. A preliminary investigation shows
that the spectra of chiral operators
of such Wolf space theories are quite
distinct from the Kaluza-Klein spectrum (\ref{KKspec}).
Another natural series of $\ca_{\gamma}$ theories to consider
is that of orbifolds of known $\ca_{\gamma}$ theories. In the case
of AdS/CFT dualities involving the small $N=4$ algebra, the
conformal field theory is a sigma model on a symmetric 
product ${\rm Sym}^k(M_4)$ \cite{va1,vast,mast,Jan,dijk,ellgen,mamost,SW}. 
Symmetric products
appear naturally as the configuration space of unordered branes,
so here we are led to consider $S_N$ orbifolds of $\ca_{\gamma}$
theories. In fact, the KK spectrum in (\ref{KKspec}) is 
highly reminiscent of that of an orbifold theory. However, 
the $\ca_{\gamma}$ theories depend on two integers $k^{\pm}$, and
the central charge in general is fractional, so it is not 
a priori clear what type of orbifold to write down for general $k^{\pm}$.
The KK spectrum looks like that of a sigma model on a
space of the form ${\rm Sym}^{k^+}({\rm Sym}^{k^-}(M))$, but
clearly this cannot be true. To get a clue one can first consider 
various limiting cases. In the limit $k^- \gg k^+$,
one of the two D1-D5 (or M2-M5 in the M-theory case) systems
approximately decouples and one expects the considerations
of the standard D1-D5 system to apply. In our case, the D5 brane
is wrapped on $S^1 \times S^3$. This suggest that the theory is
related to a sigma model on Sym${}^{k^-}(U(2))$, where we 
used the fact that $U(2)$ is topologically 
$S^1 \times S^3$. As we will see in section~7, classical
sigma models with $\ca_\g$ symmetry have
as target space a hyperk\"{a}hler manifold with torsion and 
compatible $U(2)$ action. Clearly, Sym${}^{k^-}(U(2))$ is hyperk\"{a}hler,
and has a natural $U(2)$ action. The fact that we need torsion 
implies that we should consider a $U(2)$ WZW model. This leads 
to the proposal:

For general $k^+, k^-$ 
the SCFT is a supersymmetric
sigma model with target space Sym${}^{k^-}(U(2))$,
with $k/k^-$ units of $H=dB$ flux and certain discrete
``gauge fields'' associated to the permutation group
$S_{k^-}$ turned on. 

Since the original brane configuration is invariant under
exchange of $k^+$ and $k^-$, this proposal requires the 
equivalence of the respective sigma models on 
Sym${}^{k^-}(U(2))$ and Sym${}^{k^+}(U(2))$.
As explained in section~7,
only one of the two sigma models can be weakly coupled
for given $k^+,k^-$, and in this sense this duality
is reminiscent of level-rank duality. It would be interesting
to find a direct proof of this duality.

It may seem puzzling that we allow a fractional flux of $H=dB$ in
the conformal field theory. A tentative description of this theory is
as follows. In order to define a sigma model with
target space $M$ and $H\neq 0$, 
we include a term 
\be \label{wz} S_{wz} = 2\pi i \int_{X,\partial X=
\Sigma} H
\ee
in the action. Here, $X$ is a three-manifold in target
space whose boundary is the world-sheet $\Sigma$. This term should
be independent of the choice of $X$ and this requires that $H \in
H^3(M,{\bf Z})$. It is well-known that this statement is modified
if there are also chiral world-sheet fermions
coupled to background gauge fields $A$ in the action. In this case
it is $H'=dB-t{\rm CS}(A)$ rather
than $H$ which should be in $H^3(M,{\bf Z})$, where ${\rm CS}(A)$
refers to the Chern-Simons three-form, and $t$ is proportional
to the number of chiral fermions. In our case there are no
continuous gauge fields $A$, but there are discrete $S_{k^-}$
gauge fields. By this we mean that every world-sheet
embedded in the non-singular part of Sym${}^{k^-}(U(2))$ 
naturally carries an $S_{k^-}$ bundle. Given a three-manifold $X$
with boundary $\Sigma$ and some $S_{k^-}$ bundle, a topological action
can be defined as in \cite{DW}. The different topological 
actions are classified by elements $\alpha$ of $H^3(BS_{k^-},U(1))$,
where $BS_{k^-}$ is the classifying space of the finite group
$S_{k^-}$.
We choose an element $\alpha$ such that the action for $S^3/{\bf
Z}_{k^-}$ (with the obvious ${\bf Z}_{k^-}$ bundle which is also
an $S_{k^-}$ bundle) is equal to $\exp(-2\pi i/k^-)$.
(We have not shown that
this is possible for general $k^-$, but for small $k^-$ 
we verified that such an element exists). We now include in
the path integral the topological action associated to $k\alpha$.
This depends only on $k\, {\rm mod}\, k^-$. In particular, if
$k$ is a multiple of $k^-$, this topological action is trivial.
The topological action together with the standard sigma model action
with the Wess-Zumino
term (\ref{wz}) with fractional $dB$ flux yields a 
sigma model 
that does not depend on the choice of the three-manifold
$X$, and is well-defined for all values of $k^+,k^-$.

The level of the $SU(2)$ current algebra associated to the diagonal
$SU(2)$ action on Sym${}^{k^-}(U(2))$ is given by the integral of $H$
over the orbit of $SU(2)=S^3$ in the symmetric product. This 
$S^3$ is in homology equal to $k^-$ times a 3-cycle of the
form\footnote{To see this one can e.g. look at the orbit 
of $SU(2)$ through $(1,g_0,\ldots,g_0^{k^- -1})$ where
$g_0 = {\rm diag}(e^{2\pi i/k^-},e^{-2\pi i/k^-}) \in U(2)$. 
After modding out by $S_{k^-}$ this orbit is a $k^-$-fold
cover of $SU(2)/{\bf Z}_{k^-}$, where ${\bf Z}_{k^-}$ is
generated by $g_0$.}
 $S^3/{\bf Z}_{k^-}$. It is with respect to this latter cycle
that units of flux are defined, so the level of $SU(2)$
is $k^- (k/k^-)=k$.

Conformal invariance of the sigma model will impose
constraints on the metric on Sym${}^{k^-}(U(2))$. 
For special values of $k$ there exists an exact SCFT
description of the sigma model, and our proposal 
coincides with the
conjecture in \cite{Giveonetal,Giv}. Namely, if $k$ is a 
multiple of $k^-$, $k=qk^-$, the  topological action is trivial
and the sigma model can be
described by the orbifold conformal field theory 
$(U(2)_q)^{k^-}/S_{k^-}$, where $U(2)_q$ is the
level $q$, $U(2)$, $N=1$ super WZW model. 
This is indeed a sigma model with target space
Sym${}^{k^-}(U(2))$ and $q$ units of 3-form flux. 
We expect the sigma model
for arbitrary rational $q$ to be closely related to an analytic
continuation of the exact SCFT's that exist for integer $q$.
Indeed, if we analytically continue the central charge
of $(U(2)_q)^{k^-}/S_{k^-}$ to rational $q$, we find
the right value
\be
c = k^- (3/2 +  3(q-2)/q + 3/2) = 6 k^- (q-1)/q = 6 k^+ k^- /k  
\ee
where the first factor of $3/2$ comes from the supersymmetric $U(1)$ 
factor, the second factor from the bosonic part of the $SU(2)$ WZW 
theory and the last one from the corresponding fermions. 
The truncation of spins can also be analytically continued 
to arbitrary $q=k/k^-$. In the $SU(2)_{q-2}$ theory, the maximal
spin of a primary field is $(q-2)/2$. Thus in the orbifold
the maximal spin is $k^-(q-2)/2$. This is half-integer precisely
when $q$ is a multiple of $1/k^-$.

In the remaining of this section we analyze the
spectrum of chiral primaries for some special
cases with $k=qk^-$ and compare the result to
the KK spectrum obtained from supergravity.
The conformal weights of nonchiral operators can 
change as we vary the moduli,
only the conformal weights of chiral operators are protected,
and the chiral operators with a small conformal weight
should be present in the KK spectrum in order for the
duality to be valid. 

First we consider some general aspects of $S_N$ orbifolds of
$\ca_{\gamma}$ theories. 
The RR sector of an $S_N$ orbifold can be 
decomposed in various subsectors as in \cite{dmvv}. Consider a chiral
RR state in the $Z_{p_1}$ twisted sector 
with quantum numbers $l_1^+,l_1^-$, and another one in the
$Z_{p_2}$ twisted sector with quantum numbers $l_2^+,l_2^-$. We suppress
the dependence on $U(1)$ momenta in this discussion. The conformal
weight of chiral operators in the R sector is given by
\be \label{rbound}
h_{R}(l^+,l^-,k^+,k^-)=\frac{1}{k} ( (l^+ + l^-)^2 + \frac{1}{4} k^+
  k^- ) .
\ee
Thus the two states mentioned above have conformal weight 
$h_{R}(l_1^+,l_1^-,p_1 k^+,p_1 k^-)$ and 
$h_{R}(l_2^+,l_2^-,p_2 k^+,p_2 k^-)$, if $k^+$ and $k^-$ are the
levels of the original theory that we are orbifolding.
If we combine the two states to make one in the $p_1+p_2$ twisted
sector, we do not always get a new chiral operator. We do
have the following inequality
\be
\sum_{i=1}^2 h_{R}(l_i^+,l_i^-,p_i k^+,p_i k^-) 
\geq h_{R}(l_1^+ + l_2^+, l_1^- + l_2^-, k^+(p_1+p_2),k^-(p_1+p_2))
\ee
but equality only holds if 
\be \label{condi}
\frac{l_1^+ + l_1^-}{p_1} = \frac{l_2^+ + l_2^-}{p_2}  .
\ee
Thus only specific states in the twisted sectors can combine
to give chiral primaries in the full theory. A second subtlety 
is that states that obey the equality (\ref{rbound}) are not
necessarily primaries of the original theory. 
Descendants that satisfy (\ref{rbound})
will give rise to chiral primaries in $Z_p$ twisted sectors for
sufficiently large $p$. 

Let us now apply this to the $S_{k^-}$
orbifold of the theory with $k^+=q-1$, $k^-=1$.
The latter theory has only massless representations
and the quantum numbers of the chiral 
primary operators in the
RR sector  
 (labeled by $(l^+,l^-;\bar{l}^+,\bar{l}^-)$) are \cite{Gunay}
\be
({j+1 \over 2},0;{j+1 \over 2},0),\quad 
({j+1 \over 2},0;{j \over 2},\frac12),\quad
({j \over 2},\frac12;{j+1 \over 2},0),
 \quad ({j \over 2},\frac12;{j \over 2},\frac12), \quad 0 \leq j \leq q-2 . 
\ee
The first and fourth state are bosonic, the second and third state
are fermionic. They satisfy
$l^+ + l^- = \bar{l}^+ + \bar{l}^- = {j+1 \over 2}$.
In case $k^-$ is prime, the only way to combine these 
chiral primary operators in order to get a chiral primary
operator in the orbifold theory is to take states with
$l^+ + l^- = \bar{l}^+ + \bar{l}^- = {j+1 \over 2}$
in the $Z_{j+1}$ twisted sector. This follows from
condition (\ref{condi}). By means of spectral flow
\cite{specflow} we find a set of NS states generated
by
\bea \label{y3}
 &&({j \over 2},{j \over 2};{j \over 2},{j \over 2})_S, \quad
({j \over 2},{j \over 2};{j+1 \over 2},{j+1 \over 2})_S, 
 \nonumber \\
&&({j+1 \over 2},{j+1 \over 2};{j \over 2},{j \over 2})_S, \quad
({j+1 \over 2},{j+1 \over 2};{j+1 \over 2},{j+1 \over 2})_S, \quad
0\leq j \leq q-2. 
\eea
All these states are present in the KK spectrum (\ref{KKspec}),
and form a natural ring because they satisfy $l^+=l^-,\bar{l}^+=\bar{l}^-$,
see section~3. Besides the states generated by (\ref{y3}), there will
in general be additional chiral primaries as explained above.
However, the spins of most of these states become very large as $k^{\pm}
\rightarrow \infty$, and we do not expect to see any sign of
them in the supergravity spectrum. This can be seen as follows.
If we combine states from twisted sectors they should satisfy
$\sum p_i = k^-$ and $(l_i^+ + l_i^-)/p_i=r$ where $r$ is some
fixed number independent of $i$. The spin of the combined state
satisfies $l^+ + l^- = r k^-$. After spectral flow the spins
in the NS sector obey $l^+ - l^- = (r-1/2)k^-$.  If we send 
$k^-$ to infinity, we need $r=1/2$ to keep $l^{\pm}$ finite.
Most of the additional states have $r\neq 1/2$ and are invisible
in supergravity. States with $r=1/2$ can be used to build 
arbitrary multiparticle states in the RR sector. The same
is true in the NS sector because $r=1/2$
implies $l^+=l^-$. All this supports the picture that the
chiral primaries carry the structure of a module over a ring,
the ring being generated by the chiral primaries with $l^+=l^-$.

Finally, we illustrate the appearance of additional chiral
primaries due to the existence of descendants that satisfy
(\ref{rbound}). 
Consider the case with $q=2$. 
The $N=1$ $U(2)_2$ WZW theory is a theory consisting
of one free boson and four free fermions. In the RR sector
of this theory there are several descendants that satisfy (\ref{rbound}).
Their left moving part is of the form
\be \label{y4}  t_{a_n a_{n-1} \ldots a_1}
Q^{a_n}_{-n}
Q^{a_{n-1}}_{-(n-1)} \ldots
Q^{a_1}_{-1} |\Omega\rangle \ee
where $|\Omega\rangle$ is a ground state in the Ramond sector, and
the tensor $t$ is chosen so the $SU(2)$ spins of the state
(\ref{y4}) are maximal. These descendants give rise to the RR chiral
primaries
\be 
({n+1 \over 2},{n \over 2};{n+1 \over 2},{n \over 2}), \,\,
({n+1 \over 2},{n \over 2};{n \over 2},{n+1 \over 2}), \,\,
({n \over 2},{n+1 \over 2};{n+1 \over 2},{n \over 2}), \,\,
({n \over 2},{n+1 \over 2};{n \over 2},{n+1 \over 2}),
\ee
in the $Z_p$ twisted sector when $p>n$.

%
%

\section{D-brane analysis}
\setcounter{equation}{0}

In the case of the IIB configuration, one can study the boundary SCFT 
using D-brane perturbation theory. Indeed, this is how the 
boundary SCFT was identified in the case of the D1-D5 system.
We consider the dynamics in the substringy regime,
so the relevant degrees of freedom are the ones that come from 
open strings stretching between the various branes.
By considering strings with boundary conditions dictated 
by the D-brane configuration one can easily obtain the 
massless degrees of freedom. $11$ strings yield 
a $2d$ vector multiplet and $5_15_1$ and $5_25_2$ 
yield $6d$ vector multiplets. $15_1$ and $15_2$ strings
yield $6d$ hypermultiplets. Finally there are $5_15_2$ 
strings that yield a single complex fermionic field,
localized in the intersection of the $5_1{-}5_2$ system.
One could study the system from the point of view of either
brane by  introducing a corresponding probe 
brane and studying its worldvolume theory.
Consider for instance the case of a probe D1 brane 
positioned in $(x_0, x_1)$. The  $15_1$ and $15_2$
strings imply that the worldvolume theory contains 
two hypermultiplets. These hypermultiplets 
interact in a non-local manner through the $5_1 5_2$ strings.
To lowest order one has the interaction coming from 
the one-loop diagram $15_1-5_15_2-5_21$.
One could obtain these interactions by integrating out
the $5_1 5_2$ strings. These interactions cannot be 
ignored even at low-energies as $5_1 5_2$ strings 
contain a massless degree of freedom.
For the same reason, integrating out 
these strings introduces a singularity in the 
worldvolume theory.
This complicates the analysis of the vacuum structure of the theory.
We will not attempt such an analysis here.
We only note that we also encountered some, perhaps related, 
non-locality in the supergravity description of the system.
There we saw that in order to compactify the $\lambda$ coordinate 
we had to identify large scales in the one D1-D5 system with 
small scales of the other D1-D5 system (see (\ref{identif})).
In both cases, in the large $k^+$ limit keeping $k^-$ fixed
(or vice versa) the non-locality goes away. In the same limit there 
is a weakly coupled sigma model realization of $\ca_\g$
as we discuss in the next section.

%

\section{$\sigma$ Model}
\setcounter{equation}{0}

\def\sint{\int d^2 z d^2 \theta \,}

In this section we describe a
natural class of $\sigma$-models with $\ca_{\gamma}$ symmetry
and some of their properties. This class includes the ones
that we proposed as duals of the string theory on
$adS^3 \times S^3 \times S^3 \times S^1$.
Previous work in this direction
includes the realizations of $\ca_{\gamma}$ theories via
Wolf spaces \cite{SeTrPr,SeTrPrSp,Pr,SeTh}, and on more 
general sigma models in \cite{Ketov}. 

It will be easiest to work in $N=1$ superspace. The $\ca_{\gamma}$
algebra can be written in $N=1$ superspace \cite{SeTh} because it
is a linear algebra, and contains then the super stress-energy
tensor of weight $3/2$, three spin-one supercurrents and 
four superfields of spin $1/2$. To realize this algebra we 
consider a generic $N=(1,1)$ sigma model
\be
S=\sint (g_{\mu\nu} + b_{\mu\nu}) D_+ X^{\mu} D_- X^{\nu}
\ee
with
\be
D_+ = \partial_{\theta} + \theta \partial, \qquad
D_- = \partial_{\bar{\theta}} + \bar{\theta} \bar{\partial} .
\ee
Suppose the action is invariant under a symmetry $\delta_{\epsilon} X^{\mu}$,
with $\epsilon$ satisfying $D_- \epsilon=0$. Then by varying the action
with an arbitrary unconstrained parameter the variation will be
of the form
\be \delta S = \sint (D_- \epsilon) (2 J) \ee
where $J$ is the Noether current of the symmetry. On-shell it
satisfies $D_- J=0$. Since we want to realize three spin one
and four spin one-half symmetries in the sigma model we write
as ansatz for the corresponding symmetries
\bea
\delta X^{\mu} & = & \epsilon J^{\mu}_{(a)\nu} D_+ X^{\nu}, \quad
 a=1,2,3
 \\
\delta X^{\mu} & = & \epsilon U_{(a)}^{\mu}, \quad a=0,1,2,3
\eea
where we have three antisymmetric tensors $J_{(a)}$ and
four vector fields $U_{(a)}$. As is well known, these transformation
are symmetries of the action if $J_{(a)}$ and $U_{(a)}$ are
covariantly constant with respect to the covariant derivative
with torsion,
\be
\nabla_{\rho}^+  J^{\mu}_{(a)\nu} = \nabla_{\rho}^+ U_{(a)}^{\mu} =0.
\ee
In that case we find corresponding Noether currents
\bea
\Sigma_a & = & -U_{(a)\mu} D_+ X^{\mu} \\
S_a & = & -\frac{1}{2} J_{(a)\mu\nu} D_+ X^{\mu} D_+ X^{\nu}
\eea
Furthermore, the stress energy tensor is
\be
T = -\frac{1}{2} G_{\mu\nu} D_+ X^{\mu} \partial X^{\nu}  -
\frac{1}{12} H_{\rho\mu\nu} D_+ X^{\rho} D_+ X^{\mu} D_+ X^{\nu}
\ee
with $H_{\mu\nu\rho} = \partial_{\mu} b_{\nu\rho} +
\partial_{\nu} b_{\rho\mu} +
\partial_{\rho} b_{\mu\nu} $; it generates the coordinate transformations
\be
\delta X^{\mu} = \epsilon \partial X^{\mu} + \frac{1}{2} D_+ \epsilon
D_+ X^{\mu}.
\ee
 
Next, we consider the Poisson brackets of these Noether currents. Ideally,
these should give us the OPE's of ${\cal A}_{\gamma}$. However, some
terms in the OPE's can come from higher order contractions between fields in
the Noether currents and can therefore not be seen in the Poisson
brackets. The parts of the OPE's that come from tree level
contractions should be correctly reproduced by the Poisson brackets,
and this implies several geometric constraints on the sigma model.
We will omit the details of this analysis. One of the interesting
equations one encounters is that the antisymmetric tensors $J_{(a)}$
have to satisfy, in the sense of matrix multiplication,
\be \label{huh}
J_{(a)} \cdot J_{(b)} = -\delta_{ab} +  \frac{k^--k^+}{k} \epsilon_{abc}
 J_{(c)} .
\ee
However, the algebra of $J$'s is not associative unless
\be \label{huh2}
\frac{ (k^- - k^+)^2}{k^2} =1
\ee
which means $k^+ = 0 $ or $k^-=0$. This is clearly not what we want. The
point is that (\ref{huh2}) only needs to be valid up to higher order
corrections. We can therefore allow the situation 
where $k^+ \gg k^-$ and 
$k^-/k^+$ is what controls the quantum corrections (or $k^- \gg k^+$ 
and $k^+/k^-$ controls the quantum corrections), because then (\ref{huh2})
is valid up to quantum corrections. This is also consistent with
the fact that for a sigma model interpretation the central charge
should be $3/2$ times the dimension of the target space up to
quantum corrections. Indeed, for $k^+ \gg k^-$, the central charge
is $c=6k^-(1 + {\cal O}(k^-/k^+))$, which shows that the target space
should have dimension $4k^-$, and similarly with the roles
of $k^+$ and $k^-$ reversed. An analogous situation would appear if
we would consider a WZW model for a group of dimension $4k^-$ at
level $k^+$. This is all consistent with the realizations related
to Wolf spaces, where one of the levels is related to the
the dimension of the Wolf space, and the other to the level
of the underlying WZW model.

Altogether this suggests that for $k^+ \gg k^-$, $\ca_{\gamma}$ 
theories can potentially be described by weakly coupled 
sigma models on spaces
of dimension $4k^-$. As we decrease $k^+$, the quantum corrections
become stronger and stronger, until they are so large that the
target space is no longer visible. As we then continue parameters
to $k^+ \ll k^-$, another weakly coupled sigma model description appears,
namely one where the target space has dimension $4k^+$.

Thus, a necessary condition for
the sigma model under consideration to have a $\ca_{\gamma}$ symmetry is
that the
corresponding  classical sigma model should have an $\ca_{\gamma}$ symmetry
with $k^+=0$ or $k^-=0$. The geometric conditions for this are
that the target space should be a hyperk\"ahler manifold with torsion
with compatible $U(2)$ action (as defined in \cite{OpPa}).
This should be true for both choices
of torsion in the covariant derivatives $\nabla^{\pm}\sim
\partial + \Gamma \pm H$.  It was shown in \cite{OpPa} that if $M$ is 
hyperk\"ahler with torsion and compatible $U(2)$ action, $M/U(2)$ is
quaternionic k\"ahler with torsion and all quaternionic k\"ahler manifolds
can be obtained this way. We therefore see a close relation between
quaternionic manifolds and the $\ca_{\gamma}$ algebra, but to actually
construct a sigma model we should not use the quaternionic space as
target space but rather the hyperk\"ahler space it descended from.

A space which satisfies the above requirements is the sigma model
with target space $(U(2))^{k^-}/S_{k^-}$.
This has an obvious (diagonal) action of $U(2)$ from the left
and right; once we choose a nonzero 3-form flux together with 
suitable discrete
gauge fields and appropriate
metric for this diagonal $U(2)$, the $U(2)$ action is compatible
with the hyperk\"ahler structure with torsion. Thus our candidate SCFT
has the right properties to correspond to an exact conformal
invariant $\ca_{\gamma}$ theory.

It would be interesting to understand the sigma models with 
$\ca_{\gamma}$ symmetry in some
more detail, and in particular to find the geometric interpretation
of the notion of chiral primaries and of the ring structure of
chiral primaries with $l^+=l^-$. Quaternionic manifolds have 
various interesting cohomologies \cite{coh1,coh2}, whose relation
to the chiral operators remains to be explored.

%
%

\section{Discussion and Open Problems}
\setcounter{equation}{0}

In this paper we studied the AdS/CFT duality
in the case the boundary SCFT has the large $N=4$ superconformal symmetry. 
We have presented supergravity solutions that in their 
near-horizon limit contain $AdS_3 \xx S^3 \xx S^3$, 
and computed the corresponding KK spectrum.
We proposed that the boundary 
SCFT is (possibly a deformation of) a sigma model with target 
space Sym${}^{k^-}(U(2))$, $k/k^-$ units of 3-form flux, and
suitable discrete gauge fields,
and found that this is consistent with the KK spectrum 
and has the right qualitative properties.

There are many issues that deserve further study. 
It will be interesting to analyze in more detail
the IIB configuration using D-brane perturbation theory.
In particular, to consider a probe brane and 
to obtain all couplings to the first non-trivial 
order. As noted in section 6, certain couplings 
are expected to first appear at one-loop level. 
Since there are massless states running in these 
loops, one cannot ignore these couplings 
even in low energies. These couplings, however, are of order $1/N$.
Knowing in detail the worldvolume theory will presumably also
help us understand the meaning of the identification (\ref{identif})
needed in supergravity in order to compactify one of the 
``near-horizon'' coordinates. Furthermore, the moduli 
space of the gauge theory should be related to
Sym${}^{k^-}(U(2))$.

Another issue is to provide a more precise formulation of
the boundary SCFT, and more stringent tests of
the conjectured duality. In particular, the precise meaning
of the discrete fluxes and the correct treatment of the
singularities (for a recent discussion see \cite{SW})
deserves further clarification.
A more detailed comparison of 
the spectrum of the
boundary SCFT with that of supergravity is also desirable. 
This requires
a more detailed understanding of the spectrum of
the boundary SCFT. A useful tool may be the index 
recently proposed in \cite{mamost} which plays the role
of the elliptic genus for $\ca_\g$ theories. 
Once the spectra are better understood one could go
on and compare correlation functions in the two
descriptions. 

In summary, it appears that for generic $k^+, k^-$ 
there is no useful semi-classical description. 
The $\ca_\g$ $\s$-models are only weakly coupled 
for $k^+ \gg k^-$ (or vice versa), the 
worldvolume gauge theory contains non-local 
interactions that are only suppressed 
at large $N$, and the supergravity 
solution implies a $3d$ solution only if there 
are identifications of order $1/N$.
Nevertheless, the large amount of symmetry and
exact knowledge of the theory at $k=qk^-$ may be 
sufficient to provide a non-perturbative solution.

Independently of the considerations regarding the AdS/CFT duality, 
it is an interesting question to investigate
which theories flow in the infrared to a 
fixed point with $\ca_\g$ symmetry. It is also 
interesting to further study the $\s$-models with $\ca_\g$ symmetry,
and in particular to understand the geometric notion
of chiral primaries. 

There is an interesting connection between 
our considerations and the 
issue of creation of branes when branes cross each
other. Consider the case of two D5 branes
intersecting over a string, 
say in $12345$ and $16789$, respectively, 
and with an electric field along the string. After T-duality along 
$x_1$ one gets two $D4$ branes with a relative 
velocity in the circle direction. As shown in \cite{BDG},
an open string is created when the two $D4$ 
branes cross each other. This phenomenon is 
U-dual to the creation of a single D3-brane 
when a D5-brane crosses an NS5-brane\cite{HW}.
Lifting the IIA configuration to M-theory 
we get that an M2-brane is created when two
M5 branes cross each other. In particular,
the branes are positioned as in the M-theory 
configuration we described in section 2.
Thus, the fivebranes interact through the
creation (or annihilation) of M2 branes,
and the boundary SCFT should capture 
some of the dynamics of this phenomenon.

%
%

\section*{Acknowledgements}

We would like to thank Amit Giveon, Erik Verlinde, and
especially Robbert Dijkgraaf for useful discussions. 
A.P.\ is supported by a fellowship of the Onderzoeksfonds K.U.Leuven.
This work is partially supported by the European Commission TMR
programme ERBFMRX-CT96-0045 in which 
J.  de B. and K.S. are associated to the University of Utrecht and
A.P.\ is associated to the Institute
for Theoretical Physics, K.U.\ Leuven.
A.P.\ would like to thank CERN for its hospitality while part of this
work was carried out. K.S. is supported by the Netherlands Organization 
for Scientific Research (NWO).

%
%

\appendix{$\ca_\gamma$ and $\tilde{\ca}_\gamma$ conventions}

The OPE of the $\ca_\gamma$ algebra generators 
are~\footnote{Those with the Virasoro
algebra generators are as usual.}
\bea
G_a(z) G_b(w) &=& {2c/3 \delta_{ab}\over (z-w)^3} + {2M_{ab}(w) \over
(z-w)^2} + {2L(w) \delta_{ab} + \partial M_{ab}(w) \over z-w } + \cdots
\nn
M_{ab} &\equiv& -{4\over k} \left[ k^- \alpha^{+i}_{ab} A^+_i + k^+
\alpha^{-i}_{ab} A^-_i \right]\nn
A^{\pm i}(z) G_a(w) &=& \alpha^{\pm i ~~ b}_{~~~a} \left({G_b(w)\over
z-w} \mp {2 k^\pm Q_b (w)\over k(z-w)^2} \right) + \cdots\nn
A^{\pm i}(z)A^{\pm j}(w) &=& {\epsilon^{ijk} A^{\pm}_k(w)\over z-w}
-{k^\pm \delta^{ij}\over 2(z-w)^2} + \cdots \nn
Q_a(z) G_b(w) &=& { 2\left( \alpha^{+ i}_{ab} A^+_i(w) - 
\alpha^{- i}_{ab} A^-_i(w) \right) + \delta_{ab} U(w) \over (z-w)} + \cdots\nn
A^{\pm i}(z)Q_a(w) &=& {\alpha^{\pm i~~ b}_{~~~a} Q_b(w)\over z-w}
+\cdots\nn
U(z) G_a(w) &=& {Q_a(w)\over (z-w)^2} +\cdots\nn
Q_a(z) Q_b(w) &=& -{k\delta_{ab}\over 2(z-w)} +\cdots \nn
U(z) U(w) &=& -{k\over 2(z-w)^2} +\cdots
\eea
where in complex notations $i=\{+,-,3\}$, $a=\{+,-,+K,-K\}$ and the
non-vanishing values (up to symmetry) of the various symbols are
\bea
&& \delta_{+-} = \delta_{+K\, -K}={1\over 2} \quad\qquad 
\epsilon^{+-}_{~~3}=-2i
\qquad\qquad\quad \epsilon^{3\pm}_{~~\pm}= \mp i \nn
&& \alpha^{\pm 3}_{+ -} = -{i\over 4} \qquad\qquad\qquad 
\alpha^{\pm 3}_{+K\, -K} = \mp{i\over 4} \qquad\qquad 
\alpha^{+ -}_{+\, +K} = {i\over 2} \nn
&& \alpha^{++}_{-\, -K} = -{i\over 2} \qquad\qquad\qquad 
\alpha^{- +}_{-\, +K} = - {i\over 2} \qquad\qquad 
\alpha^{- -}_{+\, -K} = {i\over 2} 
\eea
$\epsilon^{abcd}$ can be defined by
\be
\alpha^{\pm i ab} \alpha^{\pm~~cd}_{~~i} = {1\over 4} (\delta^{ac}
\delta^{bd} - \delta^{ad} \delta^{bc} \pm \epsilon^{abcd})
\ee
For the $\tilde{\ca}_\gamma$ algebra, the non trivial OPEs are
\bea
\tilde{A}^{\pm i}(z) \tilde{G}_a(w) &=& {\alpha^{\pm i ~~ b}_{~~~a} 
\tilde{G}_b(w)\over z-w} + \cdots\\
\tilde{A}^{\pm i}(z)\tilde{A}^{\pm j}(w) &=& 
{\epsilon^{ijk} \tilde{A}^{\pm}_k(w)\over z-w}
-{\tilde{k}^\pm \delta^{ij}\over 2(z-w)^2} + \cdots \nn
\tilde{G}_a(z) \tilde{G}_b(w) &=& {4\tilde{k}^+ \tilde{k}^- \delta_{ab}
\over k (z-w)^3} + {2\tilde{L}(w) \delta_{ab}\over (z-w)} 
-{8(\tilde{k}^- \alpha^{+i}_{ab} \tilde{A}^+_i(w) + \tilde{k}^+
\alpha^{-i}_{ab} \tilde{A}^-_i(w) ) \over k(z-w)^2} \nn
&& -{4\partial(\tilde{k}^- \alpha^{+i}_{ab} \tilde{A}^+_i(w) + \tilde{k}^+
\alpha^{-i}_{ab} \tilde{A}^-_i(w) ) \over k(z-w)} \nn
&& - {8(\alpha^{+i}\tilde{A}^+_i(w) - \alpha^{-i}\tilde{A}^-_i(w))_{c(a} 
(\alpha^{+j}\tilde{A}^+_j(w) - \alpha^{-j}\tilde{A}^-_j(w))_{b)}^c \over
k(z-w)} +\cdots\nonumber
\eea

%
%
%

%

%
%
%

\end{document}